\documentclass[a4paper,fleqn,usenatbib]{mnras}

\usepackage{graphicx}
\usepackage[space]{grffile}
\usepackage{latexsym}
\usepackage{amsfonts,amsmath,amssymb}
\usepackage{url}
\usepackage[utf8]{inputenc}
\usepackage{fancyref}
\usepackage{setspace}
\usepackage{flafter} 
\usepackage{xcolor}
\usepackage{soul}
\usepackage{float}
\usepackage{adjustbox}
\usepackage{stfloats}

\usepackage{newtxtext,newtxmath}

\usepackage[T1]{fontenc}
\usepackage{ae,aecompl}

\usepackage{graphicx}	
\usepackage{amsmath}	
\usepackage{amssymb}
\usepackage{makecell}
\usepackage{pbox}

\usepackage{subcaption}
\usepackage{float}
\usepackage{pdflscape} 

\usepackage{deluxetable}

\title[Orbit Retrieval using Planetary Astrometry \& Photometry]{Combining Photometry and Astrometry to Improve Orbit Retrieval of Directly Imaged Exoplanets}

\author[Bruna et al.]{Margaret Bruna,$^{1,2}$\thanks{E-mail: margaret.bruna@mail.mcgill.ca}
Nicolas B. Cowan,$^{1,2}$
Julia Sheffler,$^{3}$
Hal M.\ Haggard,$^{3}$
\newauthor
Audrey Bourdon,$^{1}$
Mathilde Mâlin$^{4}$
\\
$^{1}$Department of Physics, McGill University, 3600 rue University, Montréal, QC, H3A 2T8, CAN\\
$^{2}$McGill Space Institute, 3550 rue University, Montréal, QC, H3A 2A7, CAN\\
$^{3}$Department of Physics, Bard College, 30 Campus Rd, Annandale-On-Hudson NY 12504, USA\\
$^{4}$LESIA - Observatoire de Paris, Section de Meudon 5, place Jules Janssen, 92195 MEUDON Cedex \\
}

\date{Accepted XXX. Received YYY; in original form ZZZ}

\pubyear{2022}

\begin{document}
\label{firstpage}
\pagerange{\pageref{firstpage}--\pageref{lastpage}}
\maketitle

\begin{abstract}

Future missions like Roman, HabEx, and LUVOIR will directly image exoplanets in reﬂected light. While current near-infrared direct imaging searches are only sensitive to young, self-luminous planets whose brightness is independent of their orbital phase, reﬂected light direct imaging will reveal changes in planet brightness throughout an orbit due to phase variations. One of the ﬁrst objectives will be determining the planet’s orbit via astrometry, the projected position of the planet with respect to its host star in the sky plane. We show that phase variations can signiﬁcantly improve the accuracy and precision of orbital retrieval with two or three direct images. This would speed up the classiﬁcation of exoplanets and improve the eﬃciency of subsequent spectroscopic characterization. We develop a forward model to generate synthetic observations of the two-dimensional astrometry and the planet/star ﬂux ratio. Synthetic data are ﬁtted with Keplerian orbits and Henyey-Greenstein phase variations to retrieve orbital and phase parameters. For astrometric uncertainties of 0.01 AU in projected separation and ﬂux ratio uncertainties of $10^{-12}$, using photometry in orbit retrieval improves the accuracy of semi-major axis by 47\% for two epochs and 61\% for three epochs if the phase curves have a known shape, but unknown amplitude. In a realistic scenario where phase curve shape and amplitude are a priori unknown, photometry improves accuracy by 16\% for two epochs and 50\% for three. In general, we find that if the planetary flux is measured to better than 10$\sigma$ at multiple epochs, it usefully contributes to orbit retrieval.
\end{abstract}



\section{Introduction}
Direct imaging is the most promising approach for characterizing planets orbiting in the habitable zone of Sun-like stars, and is arguably the best way to discover such planets in the first place. Future direct imaging missions like the Nancy Grace Roman Space Telescope, HabEx, and LUVOIR will be capable of detecting visible light reflected by exoplanets \citep{2019arXiv191206219T, gaudi2020habitable, roman}. The 2021 Decadal survey has identified HabEx and LUVOIR as important missions  ``positioned to make a serious attempt at searching for biosignatures on exoearth candidates'' \citep{decadal}.  One of the first properties we seek to determine when characterizing an exoplanet is its orbit, most importantly its semi-major axis, which in conjunction with its star's luminosity is the principal determinant of a planet's climate. We would like to determine the orbit with as few imaging epochs as possible \citep[e.g.,][]{2014ApJ...795..122S}.
\subsection{Orbit retrieval via planetary astrometry}
Previous efforts to retrieve orbits of directly imaged planets focused on the time-varying position of the planet in the sky plane, i.e., planetary astrometry. This is because current direct imaging efforts are primarily sensitive to thermal emission from young Jovian planets \citep{2008Sci...322.1348M,2010Natur.468.1080M,2015Sci...350...64M,2019Lagrange}. Since the planets are self-luminous, they do not exhibit phase variations and only the changing projected position of a planet betrays its orbit. 
Indeed, many researchers have used astrometry to constrain the orbits of the four directly imaged planets around HR~8799, despite their long orbits \citep[][]{hr87992012,hr87992015,hr87992016,hr8799_k2016,blunt_2017,hr87992018}. 

Studies of direct imaging in reflected light have also focused on planetary astrometry for orbit retrieval. Compared to current thermal imaging, reflected light direct imaging favours planets on shorter orbits, better constraining the retrievals. \cite{Guimond_2019} examined the optimal number, cadence, and precision of direct imaging observations required to establish the orbit of a planet. They showed that a few epochs provide useful orbital constraints, even when some epochs are non-detections. Moreover, they demonstrated that 3 equally spaced epochs at least 90 days apart are sufficient to uniquely constrain a planet's orbit. For three or more epochs, the precision on the semi-major axis is approximately the astrometric precision multiplied by the distance to the system, e.g., three epochs of 5~mas planetary astrometry of a system 10~pc away constrains the semi-major axis to approximately 5~mas $\times$ 10~pc = 0.05~AU.  These conclusions have since been independently confirmed by \cite{Horning2019} and \cite{roman2021starshade}.

\subsection{Phase variations for orbit retrieval}
Future direct imaging missions operating in the optical and near-infrared will be sensitive to scattered light from exoplanets.  
As a distant planet orbits its star, its brightness varies as we see more of less of its illuminated hemisphere \citep{1610snml.book.....G}, so-called phase variations.  
We hypothesise that a planet's time-varying photometry could help constrain its orbit with fewer epochs of direct imaging, and significantly reduce the uncertainty on retrieved orbital parameters given the same number of epochs. 

\begin{table*}
    \centering
    \begin{tabular}{c|c|c|c}
        \hline
        \hline
         Parameter & Symbol & Input Distribution & Prior Distribution  \\
         \hline
         Semi-major axis & $\ln a$ & $a = 1$~AU & $\mathcal{U}[\ln 0.01 \mathrm{AU}, \ln 50  \mathrm{AU}]$   \\ 
         Eccentricity & $e$ & B($\alpha=0.867$, $\beta = 0.303$)& B($\alpha=0.867$, $\beta = 0.303)$ \\
         Inclination & $\cos{i}$ & $\mathcal{U}[\cos0, \cos\pi/2]$ & $\mathcal{U}[\cos0,\cos\pi]$ \\
         Argument of periapsis & $\omega$ & $\mathcal{U}[0,2\pi)$ & $\mathcal{U}[0,2\pi)$ \\
         Longitude of ascending node & $\Omega$ &  $\mathcal{U}[0,2\pi)$ & $\mathcal{U}[0,2\pi)$ \\
         Mean anomaly of first epoch & $M_{o}$ & $\mathcal{U}[0,2\pi)$ & $\mathcal{U}[0,2\pi)$ \\
         \hline
         Albedo figure of merit & $\ln AR_{p}^{2}$ & $AR_{p}^{2} = 0.3 R_{\oplus}^{2}$ & $\mathcal{U}[\ln(0.01R_{\oplus}^{2}), \ln{R_{\rm Jup}^{2}}]$\\
        Heyney-Greenstein parameter & $g$ & $g=0$ or $\mathcal{N}(0,0.7)$ & $g=0$ or $\mathcal{N}(0,0.7)$\\
        \hline
        \hline
    \end{tabular}
    \caption{\textbf{Six Keplerian parameters (top) and two phase curve parameters (bottom)}. The input distributions are used when generating synthetic planets, while the prior distributions are used when retrieving orbits.  These distributions are the same except for the semi-major axis, $a$, and the reflection figure of merit, $AR_p^2$.   The input semi-major axis, albedo and radius correspond to those of Earth so that all of our synthetic planets are Earth-like. Their priors, on the other hand, are broad and log-uniform because planets tend to be small and on close orbits.  For orbital eccentricity we use a beta distribution with $\alpha = 0.867$ and $\beta = 0.303$ \citep{nielsen_2008,kipping_2013}.  
    }
    \label{table:params}
\end{table*}

\begin{figure}
    \centering
    \includegraphics[scale=0.35]{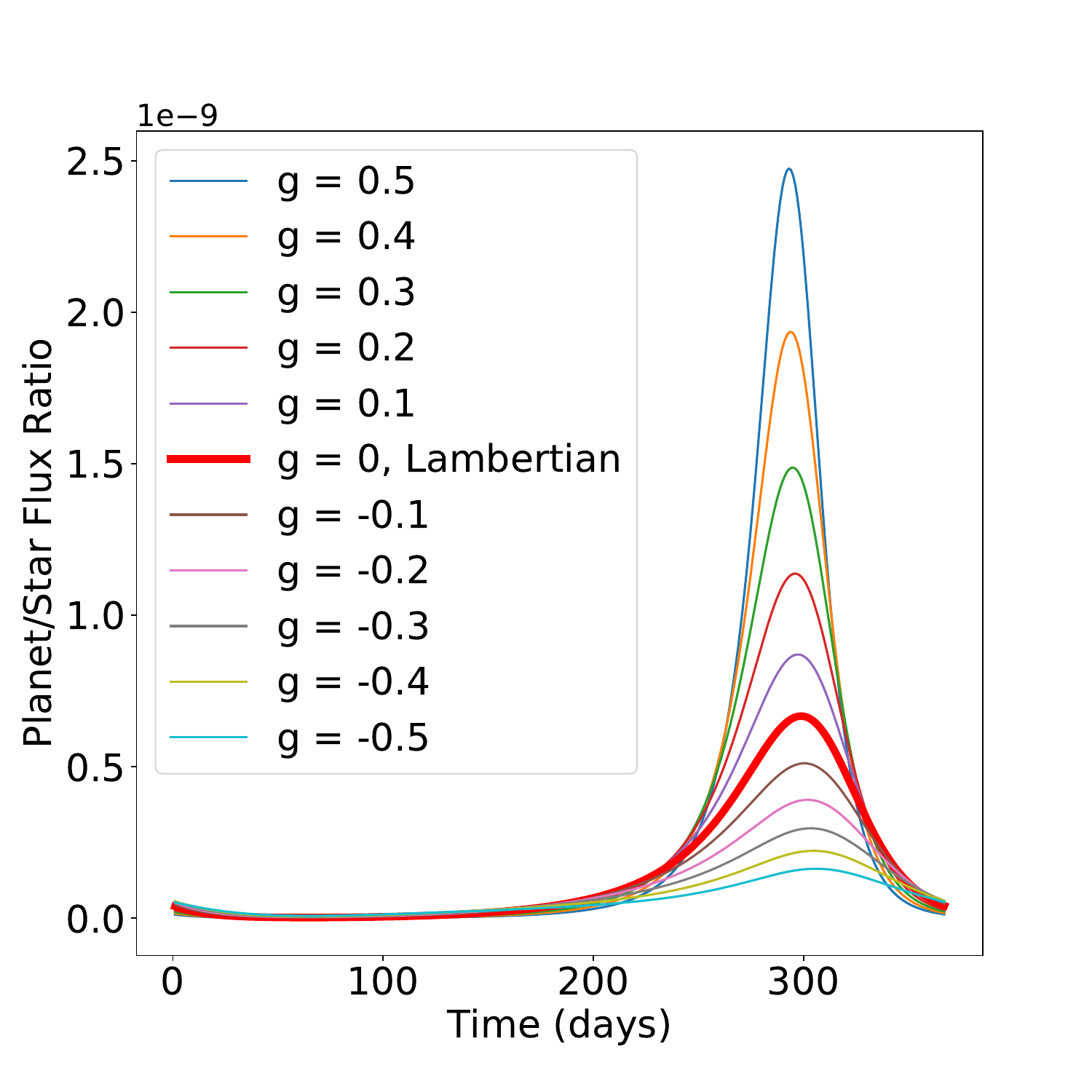}
    \caption{Brightness model produced by the Henyey-Greenstein phase function for a scattering parameter $g$ ranging from -0.5 to 0.5. The curves are generated with $a = 1.00$ AU, $e = 0.31$, $\cos i = 0.41$, $\omega = 2.25$ rad, $\Omega = 3.38$ rad, and $M_o = 3.87$ rad, and $AR_{p}^{2} = 3.63\times10^{-10}$ AU$^{2}$ (where $A = 0.3$ and $R_{p} = R_{\oplus}$).}
    \label{fig:nl_all}
\end{figure}

In Section 2 we describe our numerical experiment. Section 3 presents the results of orbital retrievals for planets detected at two, three, and four epochs and for a variety of astrometric and photometric uncertainties. We compare models that retrieve planets exibiting Lambertian phase curves and ones which have a Henyey-Greenstein phase curve. In Section 4 we discuss the impact of these results on the design of future missions.  

\section{Methods}\label{sec:numerical}

\subsection{Keplerian Orbits and Henyey-Greenstein Phase Curves}

The projected sky position of a planet moving on a Keplerian orbit is given by \citep{xy}:
\begin{equation}\label{eq:x}
    x = r\Big(\mathrm{cos}\Omega \ \mathrm{cos}(\omega + \nu) - \mathrm{sin}\Omega \ \mathrm{sin}(\omega + \nu) \ \mathrm{cos}i\Big)
\end{equation} and 
\begin{equation}\label{eq:y}
    y = r\Big(\mathrm{sin}\Omega \ \mathrm{cos}(\omega + \nu) + \mathrm{cos}\Omega \ \mathrm{sin}(\omega + \nu) \ \mathrm{cos}i\Big)
\end{equation} where the planet--star separation is
\begin{equation}\label{eq:r}
    r = \frac{a(1 - e^2)}{1 + e \cos E}.
\end{equation} 
In the above, $\Omega$ is the longitude of ascending node, $\omega$ is the argument of periapsis, $\nu$ is the time-dependent true anomaly, $i$ is the inclination, $a$ is the semi-major axis, $e$ is the eccentricity, and $E$ is the eccentric anomaly. 

Synthetic planets are generated using the parameter distributions outlined in Table \ref{table:params}. The time dependence of the true anomaly, $\nu$, is computed using Newton's method. To produce and retrieve orbits, we reparameterize our model using the mean anomaly, $M$, because it has a uniform prior over $(0,2\pi)$:
\begin{equation}\label{eq:mean_anom}
    M = E - e\sin{E}
\end{equation} where the eccentric anomaly is related to the true anomaly via
\begin{equation}\label{eq:mean_anom2}
    \tan{E} = \frac{\sqrt{1 - e^{2}}\sin{\nu}}{e + \cos{\nu}}.
\end{equation}
We adopt the mean anomaly at the first epoch, $M_0$, as our initial condition. Since mean anomaly advances at a constant rate throughout a planet's orbit, the prior on $M_0$ is uniform.

The reflected flux ratio of the planet to its host star is \citep{old_flux}: \begin{equation}\label{eq:epsilon}
    \centering
    \epsilon(\Phi) \equiv \frac{f_{p}(\Phi)}{f_{\star}} =  A_{g} P_{\rm HG}(\Phi,g)\frac{R_{p}^{2}}{r^{2}},
\end{equation} 
where $A_{\mathrm{g}}$ is  the geometric albedo, $P_{\rm HG}(\Phi,g)$ is the phase function, and $R_p$ is the planetary radius.

We adopt the phase curve parameterization of  \cite{henyey_greenstein_1941}: 
\begin{equation}\label{eq:HGL}
    P_{\rm HG}(\Phi,g) = \frac{1 - g^2}{(1 + g^2 - 2g\cos \Phi)^{\frac{3}{2}}} \left(\frac{\sin \Phi + (\pi - \Phi) \cos \Phi}{\pi} \right),
\end{equation}
where the star--planet--observer phase angle $\Phi$ is given by
\begin{equation}\label{eq:phase_angle}
    \cos \Phi = \sin(\omega + \nu) \sin i. 
\end{equation} For $g = 0$ the scattering is diffuse (i.e isotropic) \citep{Lambert1760} and $P_{\rm HG}$ reduces to the Lambertian phase curve \citep{Russell1906}. Figure~\ref{fig:nl_all} shows example HG phase variations as function of time and for different scattering parameters $g$.

\subsection{Synthetic planets and observations}\label{sec:synthetic}
We generate 100 synthetic Earth-like planets and for each one we produce synthetic astrometric and photometric data for 2, 3, and 4 epochs.  We choose this sample size because our results are unchanged if the number of synthetic planets is doubled.

We randomly generate the six Keplerian and two phase curve parameters according to the distributions outlined in Table \ref{table:params} to produce an orbit and a phase curve. We assume single planet systems for the entirety of this experiment and we enforce that each direct-imaging epoch yields a detection of the planet.  The detections-only simplification minimally impacts our results, as we discuss in \S\ref{sec:conclu}. The target stars of future direct imaging missions have well characterized masses \citep{gaudi2020habitable,2019arXiv191206219T}. We therefore do not retrieve the period of the planet independently of its semi-major axis, since the mass of the system will be almost entirely dominated by that of the host star.

All of our synthetic planets have a semi-major axis of $a=1$~AU, an albedo of $A=0.3$, and the radius of the Earth, $R_p = R_\oplus$. 
The orbital eccentricity, $e$, is drawn from a beta distribution \citep{nielsen_2008,kipping_2013}, the orbital inclinations, $i$ are isotropic (uniform in $\cos i$), while the argument of periastron, $\omega$, longitude of ascending node, $\Omega$, and mean anomaly at the first epoch, $M_0$, are drawn from uniform distributions.  

\begin{figure*}
    \centering
    \includegraphics[scale=0.45]{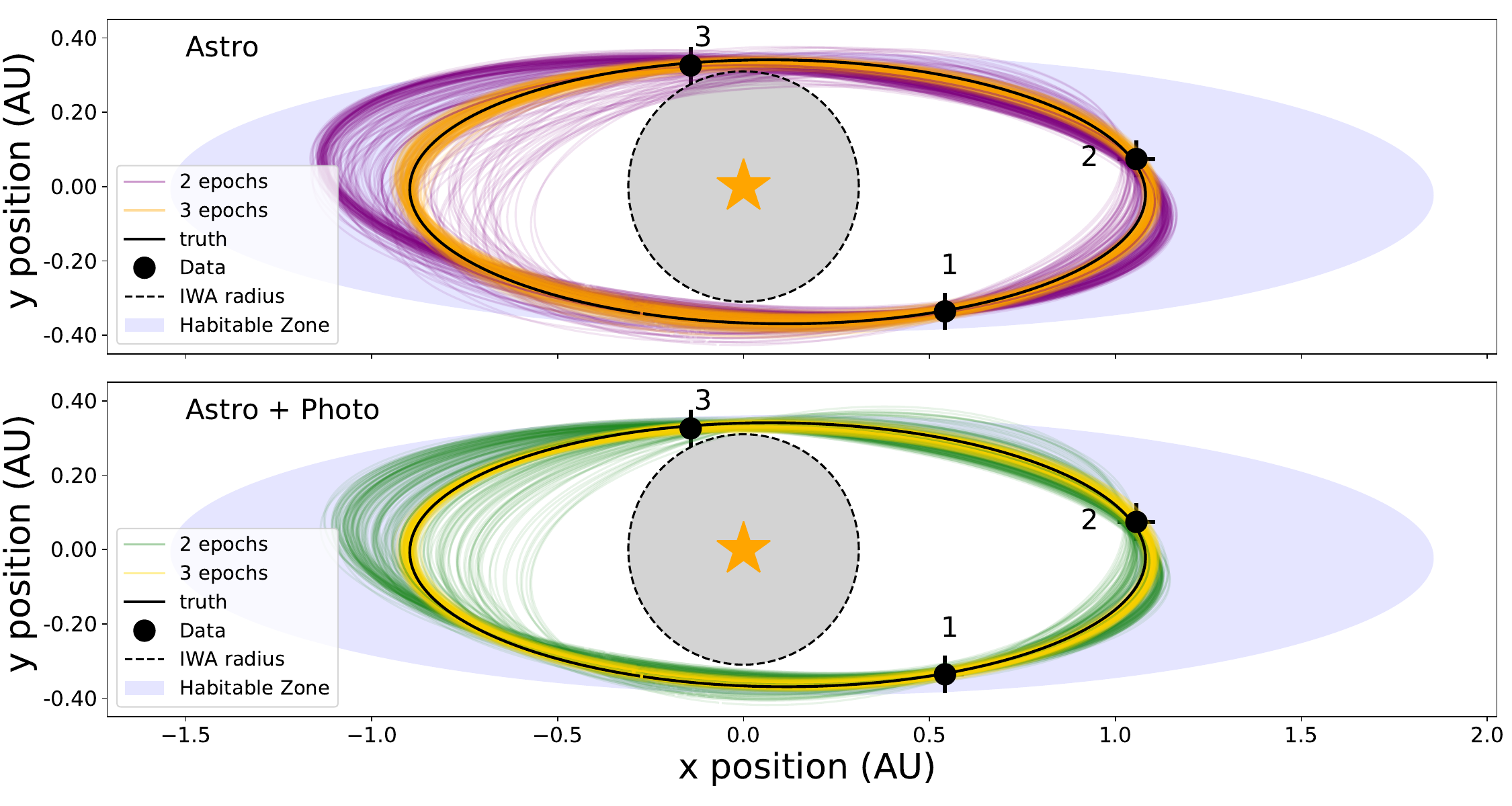}
    \caption{Orbit retrievals for an exoplanet after 2 and 3 detections. The top panel demonstrates orbital retrievals using only astrometry, and the panel below demonstrates retrievals using astrometry and photometry. The colored lines are draws from \texttt{kombine} in the 2 epoch case, and from \texttt{emcee} in the 3 epoch case. The black line is the true orbit of the synthetic planet, and the black data are the astrometry at the three epochs. The astrometric uncertainties of 0.01 AU have been inflated $5\times$ to be visible. The gray circle around the host star indicates the area obscured by a notional coronagraph or starshade. The lavender shaded region indicates the habitable zone predicted by \protect\cite{Kopparapu_2013}.}
    \label{fig:orbitretrieval}
\end{figure*}

Following \cite{Guimond_2019}, we set the distance to the star-planet system to 10 pc and adopt an inner working angle (IWA) of 30 mas, corresponding to a minimum projected separation of 0.3 AU.  
We produce images at each epoch given a fixed 90-day cadence \citep{Guimond_2019}. We add Gaussian astrometric noise of $\sigma_{\rm astro}$ to the projected $x$ and $y$ positions of the planet and $\sigma_{\rm photo}$ to the planet/star flux ratio; we also adopt these values for the astrometric and photometric measurement uncertainties.  
We show example multi-epoch observations in Figure \ref{fig:orbitretrieval}.
\subsection{Retrieving orbital and phase parameters}
Retrieving orbital and phase parameters entails performing fits to astrometric and photometric data. For astrometric fits to $N$ epochs we define the usual badness-of-fit,
\begin{equation}\label{eq:chi2_astro}
    \chi_{\rm astro}^2 = \sum_{i = 1}^N \frac{(x_i - x_{m,i})^2 + (y_i - y_{m,i})^2}{\sigma_{\rm astro}^2},
\end{equation}
where $x_i$ and $y_i$ are the measured location of the planet at the $i$th epoch while $x_{m,i}$ and $y_{m,i}$ are the model prediction for that epoch.

For retrievals using both astrometry and photometry we define the total badness-of-fit as $\chi^2 = \chi_{\rm astro}^2 + \chi_{\rm photo}^2$, where the photometric badness-of-fit is
\begin{equation}\label{eq:chi2_photo}
    \chi_{\rm photo}^2 = \sum_{i = 1}^N \frac{(\epsilon_i - \epsilon_{m,i})^2}{\sigma_{\rm photo}^2}.
\end{equation}
Here $\epsilon_i$ and $\epsilon_{m,i}$ are the measured and predicted planet-to-star flux ratio.

\subsubsection{Posterior Sampling}

We begin each retrieval by performing a $\chi^2$ minimization using \texttt{scipy.optimize} to obtain a first guess of the best fit parameters. We then use ensemble samplers \texttt{emcee} \citep{emcee} and \texttt{kombine} \citep{kombine} to retrieve the posterior distributions on orbital and phase parameters. These Bayesian codes require a likelihood function, which we define as $\ln L = -\chi^2$, and a prior probability distribution for the fitted parameters. 

With the exception of semi-major axis $a$ and the reflected light figure of merit $AR_p^2$, we adopt as priors the same distributions used to generate the synthetic planets.  For semi-major axis and $AR_p^2$ we adopt log-uniform priors to encode that there are more planets orbiting closer to stars and there are more small planets, following previous work \citep{petigura_2013,Foreman_Mackey_2014,Silburt_2015,Burke_2015,Christiansen_2015,Kopparapu_2018}.  Our priors are listed in Table~\ref{table:params}. 

For under-constrained retrievals (2 epochs) we use \texttt{kombine}, which uses a clustered kernel-density-estimate proposal that allows for more efficient sampling when plausible solutions are spread out in parameter space, as one would expect for a formally degenerate problem. We use 500 walkers that take 800 to 8000 steps, with burn-in ranging from 300 to 3000 steps; \texttt{kombine} checks for convergence automatically, so the number of steps for a particular run varies from one fit to another. 

For marginally- or over-determined problems (3 or more epochs) we use \texttt{emcee} to retrieve posterior distributions on the Keplerian and phase parameters. We use 50-100 walkers that run for 5000--15000 steps, depending on the model in question. 
For both \texttt{kombine} and \texttt{emcee} we check for convergence by examining the corner plots and walker plots produced.  As a spot check, we repeated some of the 3-epoch retrievals using \texttt{kombine} to ensure that the resulting posteriors were indistinguishable from those obtained with \texttt{emcee}.   

Figure~\ref{fig:orbitretrieval} shows an orbit retrieval for 2 and 3 epochs with and without photometry. 
The precision of our retrieval depends on the use of photometry, which adds one datum per epoch but also one or two fitted parameters, depending on the choice of phase curve parameterization.

The retrieved parameters are compared to the true values for that synthetic planet to determine the bias and accuracy of the retrievals.  The discrepancy between the retrieved parameter and its true value is denoted by $\Delta$.  For a given parameter, the mean discrepancy for a large number of synthetic planets is an estimate of the retrieval bias, $\langle\Delta\rangle$, while the standard deviation of these same discrepancies is an estimate of the retrieval accuracy $\sigma_{\Delta}$. We tabulate bias and accuracy for all fitted parameters, but we focus on the semi-major axis because it is the primary discriminant for identifying potentially habitable planets, a stated goal of next generation direct imaging missions. 

\begin{figure*}
    \centering
    \includegraphics[scale=0.35]{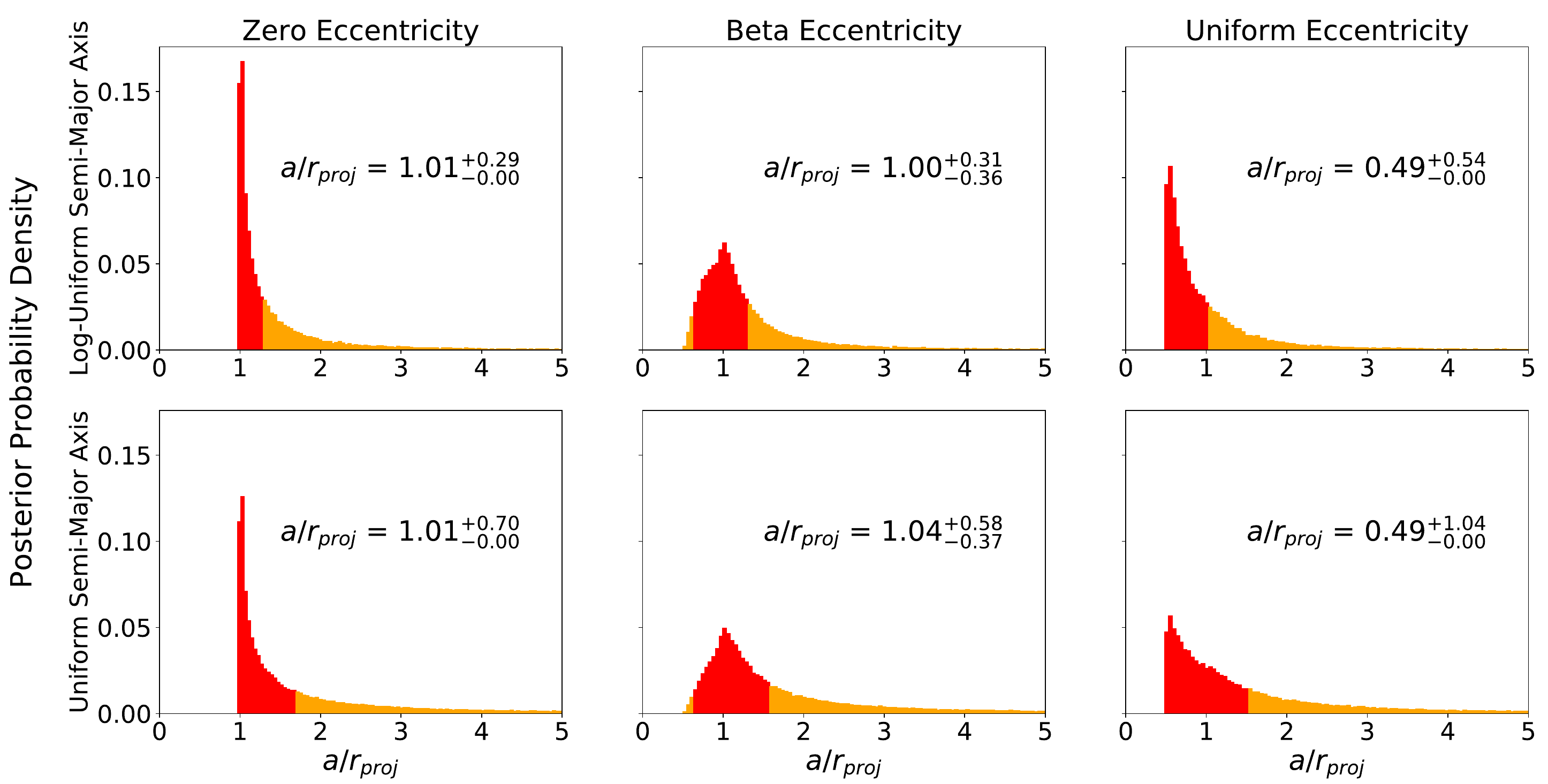}
    \caption{A single measurement of $r_{\rm proj}$ from direct imaging or planetary microlensing constrains the planet's semi-major-axis.  Each panel shows the posterior probability distribution of $a/r_{\mathrm{proj}}$  for different prior distributions on semi-major axis and eccentricity. The prior on semi-major axis is either  log-uniform (top row) or uniform (bottom row). The eccentricity is either set to 0 (left column), given a beta distribution (middle column), or uniform distribution (right column).  Concretely, if a single epoch of infinitely precise astrometry shows the planet to have a projected separation of $r_{\rm proj}=1$~AU, then the x-axis is simply the semi-major axis of the planet in AU; for imprecise astrometry, these distributions would have to be convolved with a Gaussian representing the measurement uncertainty. The peaks in the data therefore correspond to the most probable semi-major axis and the red bins show the 1$\sigma$ (68\% confidence) interval. The top-left panel has the most optimistic priors: the posterior peaks very close to the true value and has an asymmetric 1$\sigma$ interval spanning 0.29~AU. The bottom-right panel has the most pessimistic priors: the posterior is severely biased and the 1$\sigma$ interval spans 1.04~AU. The top-center panel is the most realistic case: the posterior is essentially unbiased and the 1$\sigma$ interval spans 0.67~AU.   
    } 
    \label{fig:hists}
\end{figure*}

\subsubsection{Number of epochs} 
We consider scenarios with detections of the planet at 2, 3, or 4 epochs. Purely astrometric retrievals of 2 epochs use 4 data to retrieve 6 parameters. Meanwhile, photometric + astrometric retrievals of 2 epochs fit 6 data with 7--8 parameters, depending on the assumed scattering phase function.  Thus orbit retrieval based on 2 epochs of direct imaging is always under-constrained. With 3 detections of the planet, astrometric + photometric retrieval fits 9 data to retrieve 7--8 parameters, while retrievals based solely on astrometric information fit 6 data with 6 parameters. 

With detections at 4 epochs, the orbit is over-determined regardless of whether photometry is considered, so retrieval is an optimization problem. 

\subsubsection{Astrometric and photometric uncertainty} 
We adopt fiducial uncertainties of $\sigma_{\rm astro}=0.01$~AU for astronometry and $\sigma_{\rm photo}=10^{-12}$ for photometry. Our fiducial astrometric uncertainty is smaller than that used by \cite{Guimond_2019}, \cite{roman2021starshade}, or the predicted uncertainty of 0.05 AU by \cite{gaudi2020habitable} and \cite{2019arXiv191206219T}. Photometric uncertainties of 1\% have been adopted in previous studies \citep{Cowan_2009,Fujii_2012,Farr_2018}. Our chosen astrometric and photometric uncertainties are optimistic in order to focus on the intrinsic degeneracies of the retrieval problem in the 2 epoch case.  We repeat our experiment with more realistic errors of $\sigma_{\rm astro}=0.035$ AU and $\sigma_{\rm photo}=3.5\times10^{-12}$ \citep{Guimond_2019, 2019arXiv191206219T}. To put these values in perspective, the intrinsic degeneracy in the single epoch case leads to uncertainties of $\sigma_a =0.29$--1.04~AU (see \S 3.1), while the amplitude of orbital phase variations for an Earth-twin are on the order of $10^{-10}$.

\subsubsection{Scattering function for phases}
We consider two scenarios regarding prior knowledge of the planet's scattering phase function.  In the optimistic case we have good prior knowledge: we assume the same HG parameter in the retrieval as we use to produce the synthetic phase curves ($g=0$ in both cases, i.e., Lambertian phase curves all around). In the realistic scenario, synthetic planets have randomly generated $g$ drawn from a Gaussian distribution inspired by Solar system worlds. We fit for $g$ as part of our retrieval, using the same Gaussian distribution as our prior (see Table \ref{table:params} for details).

\section{Results}\label{sec:results}

\subsection{Single epoch posterior on semi-major axis}
Before presenting the results of our orbit retrievals for multi-epoch direct-imaging campaigns, it is useful to consider the orbital information present in a single epoch. A single epoch of direct imaging \citep{Marois2008} or planetary microlensing \citep{Gould1992} provides a measurement of the planet's projected separation from its host star. The instantaneous projected separation places constraints on the planet's semi-major axis: to first order, the two are equal. In detail, the posterior distribution for semi-major axis depends on the choice of priors for semi-major axis and orbital eccentricity. 

\begin{figure*}
    \centering
    \includegraphics[scale=0.35]{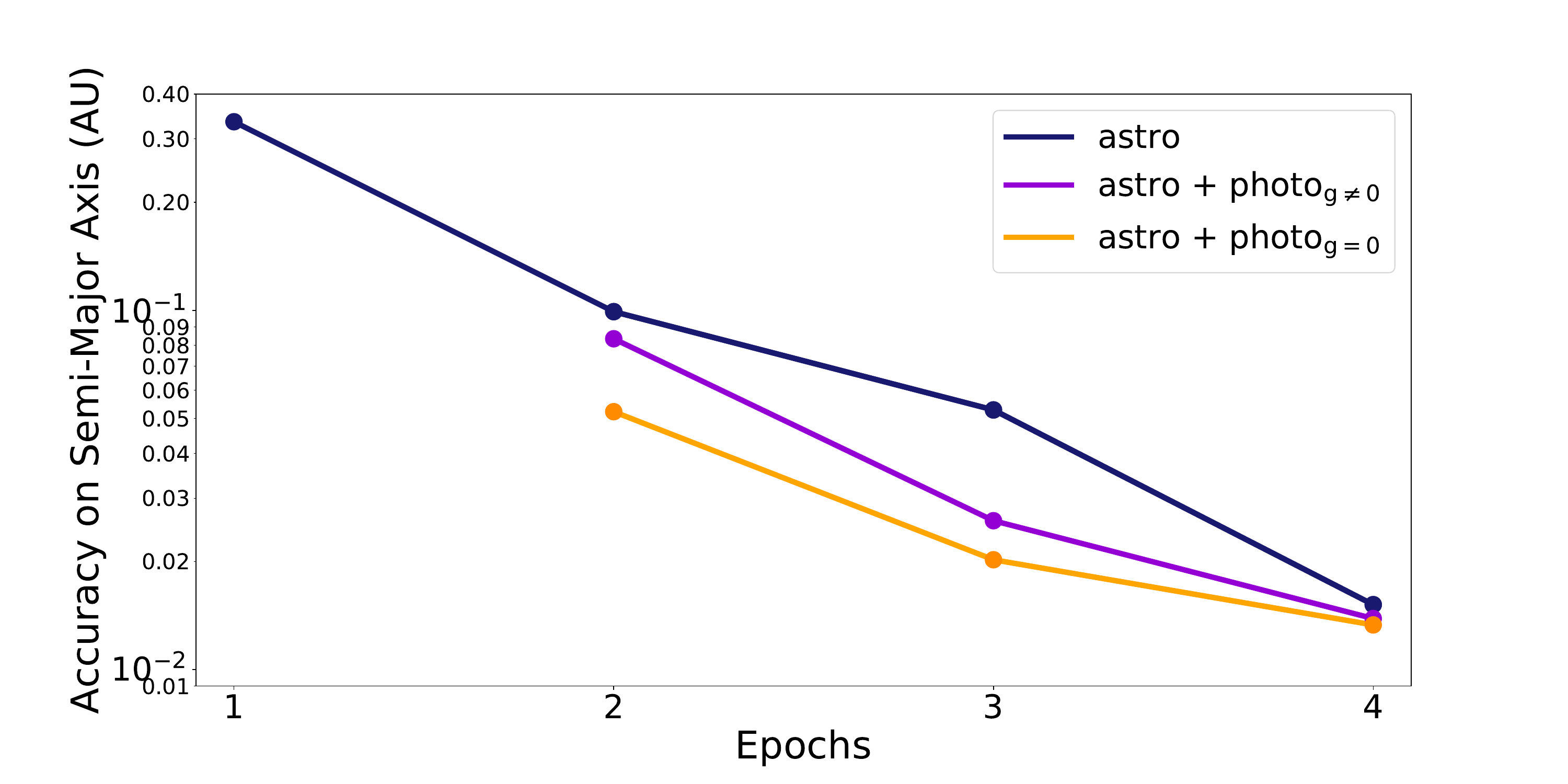}
    \caption{The 68\% confidence interval ($\sigma_\Delta$) on the accuracy of the semi-major axis for 100 runs at 1, 2, 3, and 4 detections. The blue line indicates these values for a model that only includes astrometry. The $\sigma_\delta$ for 1 epoch is taken from the the top-center panel of Figure \ref{fig:hists}. The yellow line is for a model including astrometry and photometry, where the phase curve is Lambertian and we presume to know this in our retrieval. The purple line also indicates a model with photometry, but where the generated planets have a Henyey-Greenstein phase curve, and the scattering parameter g is part of the fit. Models with astrometry + photometry provide better accuracy, particularly when the phase curve is Lambertian. Models with more epochs also yield improved accuracy, as expected.}
    \label{fig:err_plot}
\end{figure*}

We generate 5 million planets with the prior distributions shown in Table~\ref{table:params}, and retrieve those with projected separations of $r_{\rm proj}\equiv \sqrt{x^2+y^2} \in [0.99, 1.01]$~AU. Note that for single-epoch observations, the projected separation is the only useful datum: the $x$ and $y$ positions are not useful on their own, while we conservatively neglect the marginal information content in a single epoch of planetary photometry \citep{Guimond_2018,Bixel2020}. We show in Figure~\ref{fig:hists} the posterior probability densities for the scaled semi-major axis, $a/r_{\rm proj}$. The peaks in the data therefore correspond to the most probable semi-major axis and the red bins show the 1$\sigma$ (68\% confidence) interval. Concretely, if a single epoch of direct imaging shows a planet at a projected separation of $r_{\rm proj}=1$~AU, then these are simply the posterior distributions on the semi-major axis. We adopt a uniform or log-uniform prior on semi-major axis in astronomical units. This choice has a negligible impact on the posterior peak, but changes the width of the posterior as expected: log-uniform favours shorter $a$ leading to narrower distributions.

The prior on eccentricity is either a delta function fixed to 0 (for planets with circular orbits), a beta distribution (the same as outlined in Table \ref{table:params}), or uniform. These significantly impact the overall shape of the distribution and the peak of the posterior. With a uniform eccentricity the peak of the scaled semi-major axis is at 0.5 because a planet with an eccentricity of 1 will spend the majority of its orbit near apastron, which is approximately located at a distance of $r_{\rm{proj}}$ $\lesssim 2a$ away from the star. The single-epoch projected separation provides an unbiased estimate of the semi-major axis if the eccentricity is zero or described by a beta distribution, but overestimates the semi-major axis by a factor of two for uniformly-distributed eccentricities. In many other astrophysical contexts eccentricity is expected to have a thermal distribution, where the probability increases linearly from 0 to 1. \cite{star_paper} studied M dwarf binary systems and determined that, with a thermal eccentricity prior, the mean of the posterior of $a/r_{\mathrm{proj}} = 1.26$. We are able to reproduce this result with the same eccentricity distribution and a uniform prior on semi-major axis.

\begin{figure*}
    \centering
    \includegraphics[scale=0.45]{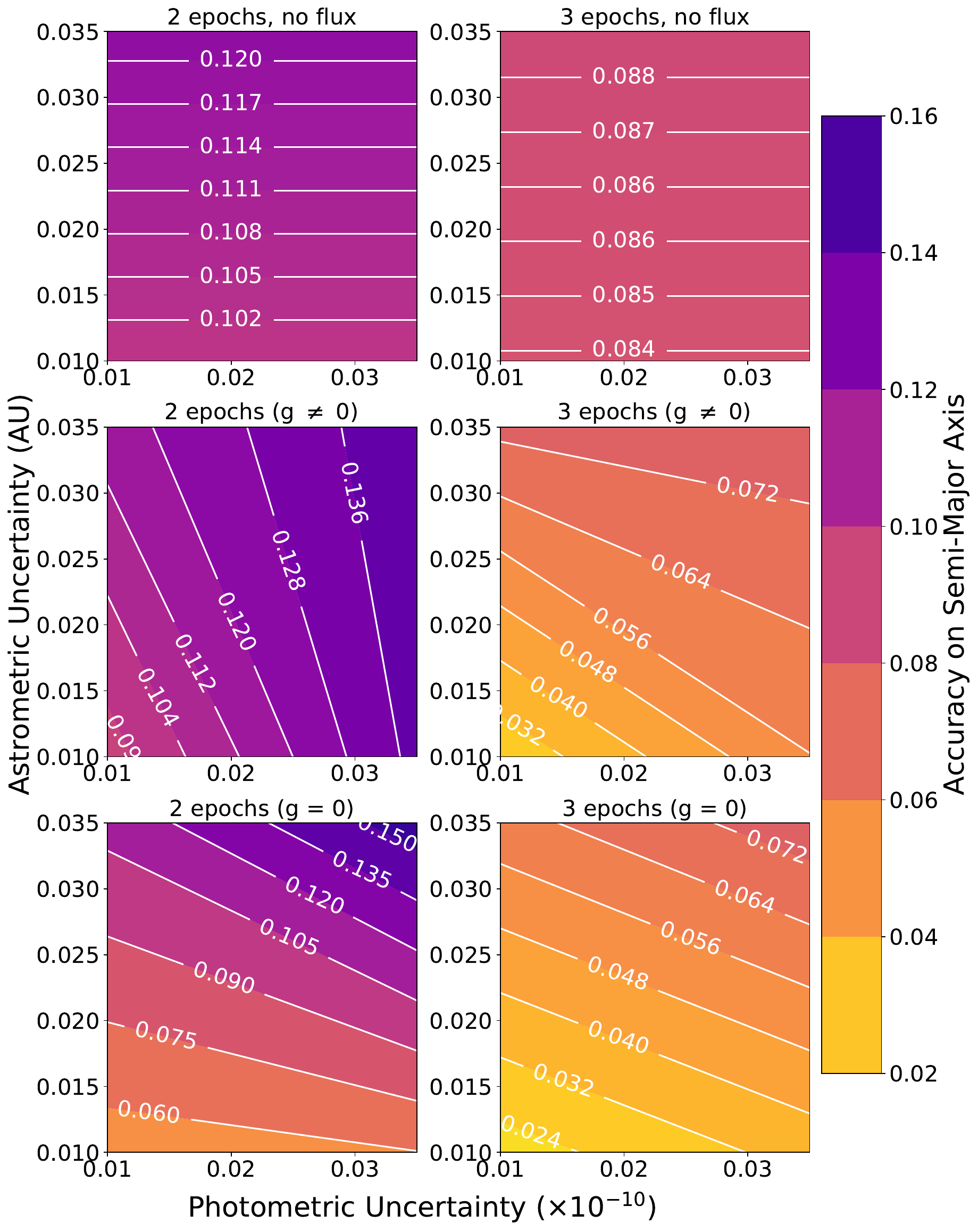}
    \caption{The accuracy of retrieved semi-major axis as a function of astrometric and photometric measurement uncertainty. Contour plots indicate the 68\% confidence interval for 100 retrieved semi-major axis values ($\sigma_{\Delta a}$) determined for varying uncertainty values on photometry and astrometry of the synthetic data, for values of 0.01 and 0.035 AU for  astrometry, and $10^{-12}$ and $3.5\times10^{-12}$. The left panels correspond to retrievals done for 2 epochs, and the right for retrievals done with 3 epochs. The top two panels indicate results for retrievals done with only astrometry (hence the constant values over increased photometric uncertainty). The middle two panels demonstrate results for retrievals done with astrometric + photometric information with a Henyey-Greenstein phase function, and the bottom two for a Lambertian phase function. Measurement uncertainty on astrometry impacts accuracy results more significantly than photometric uncertainty, except for 2 epochs of data with a Henyey-Greenstein phase function. With only 2 epochs of data, higher uncertainties on astrometry and photometry can provide worse accuracy than if no photometry is used.}
    \label{fig:3epoch_contour}
\end{figure*}

The priors on semi-major axis and orbital eccentricity both significantly affect the width of the posterior.  The most optimistic priors ($e=0$ and log-uniform $a$) result in an asymmetric 1$\sigma$ interval spanning 0.29~AU. The most pessimistic priors (uniform $e$ and $a$) produce a 1$\sigma$ interval spanning 1.04~AU. The most realistic case (beta-distributed $e$ and log-uniform $a$) results in a 1$\sigma$ interval spanning 0.67~AU; dividing this interval by 2 yields the left-most point in Figure \ref{fig:err_plot}.

Precise astrometry can be taken to mean $\sigma_{\rm astro}$ is much smaller than the intrinsic ranges shown in Figure~\ref{fig:hists}.  We therefore expect that for measurement uncertainties less than 0.3 AU, these distributions will not be much affected and the posterior distribution is approximately independent of the astrometric uncertainty. But for three or more epochs, the accuracy of the retrieved semi-major axis will be approximately proportional to the astrometric uncertainty.

\subsection{Orbit retrievals for 2, 3, and 4 epochs}
For two or more epochs of direct imaging, photometry can play a useful role in constraining the orbit of a planet. Figure \ref{fig:err_plot} compares the accuracy of semi-major axis retrievals,  $\sigma_{\Delta}$, for 2, 3, and 4 epochs. 
There is minimal improvement in the accuracy between 3 and 4 epochs, which is to be expected due to the number of data and parameters \citep{Guimond_2019}: once the retrieval problem becomes over-determined, it is an optimization problem and we expect accuracy to improve as $N^{-\frac{1}{2}}$. We focus primarily on retrievals based on 2 or 3 epochs since these are cases where photometry significantly impacts retrieval accuracy.

At 2 epochs, all models constrain the retrieved semi-major axis to within 10\%, where there is minimal improvement with the addition of a Henyey-Greenstein phase curve. Retrievals using a  Lambertian phase curve brings retrievals close to within 5\%. Notably, this is comparable to using 3 epochs of only astrometric data. At 3 epochs, using either a Lambertian or Henyey-Greenstein phase curve provides significant improvement and constrains results within 5\%.

\begin{figure*}
    \centering
    \includegraphics[scale=0.2]{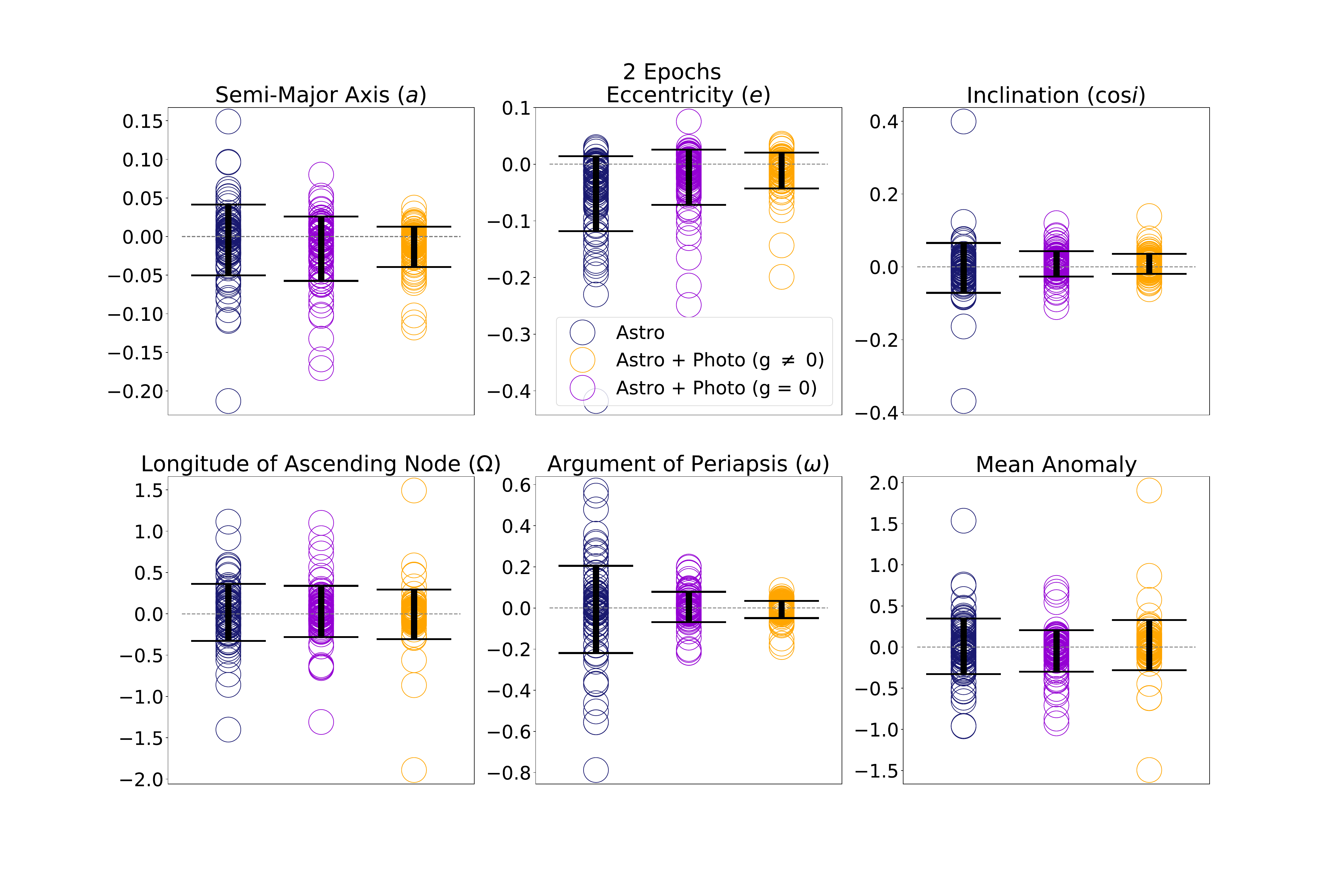}
    \includegraphics[scale=0.2]{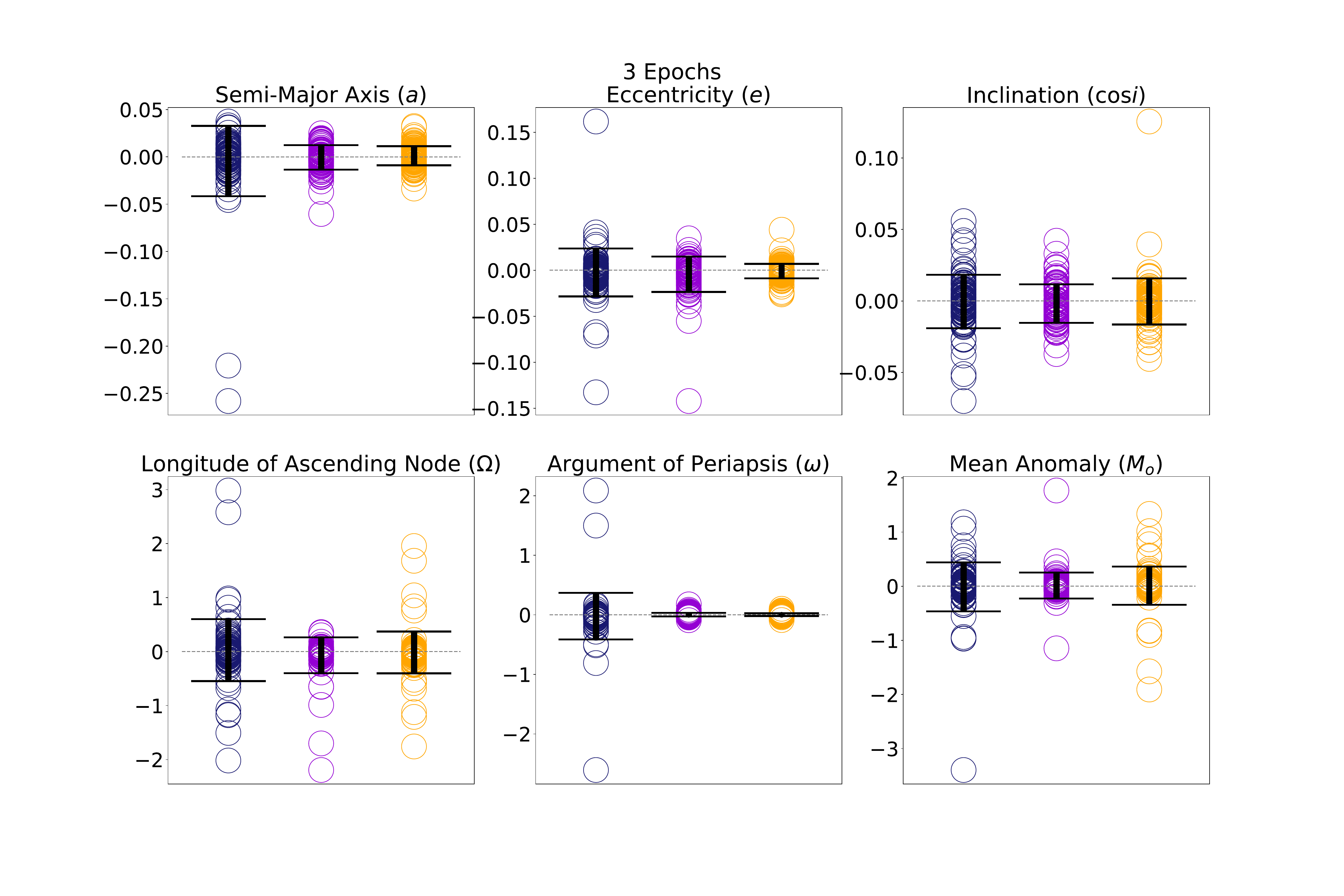}\\
    \caption{Accuracy of retrieved Keplerian parameters for two (top) and three (bottom) epochs of direct imaging 90 days apart. Each data point is the difference between the true parameter value and the one retrieved ($\Delta$) and the error bars indicate the 68\% confidence interval on these results for 100 planets  ($\sigma_{\Delta}$). Each panel corresponds to the retrievals performed for different Keplerian parameters. We demonstrate results for retrievals with astrometry (blue), astrometry + photometry with a Henyey-Greenstein phase curve (purple), and ones with astrometry + photometry with a Lambertian phase curve (yellow). These figures indicate that retrievals for the argument of periapsis ($\omega$) benefit most from added photometric information, followed by semi-major axis ($a$) and eccentricity ($e$). Retrievals for the longitude of ascending node ($\Omega$) and the mean anomaly ($M_o$) benefit the least.}
    \label{fig:allparams}
\end{figure*}

Figure \ref{fig:3epoch_contour} compares the 1$\sigma$ accuracy of retrieved semi-major axis ($\sigma_\Delta$) as a function of measurement uncertanties for 2 and 3 epochs. If the posterior distributions on the retrieved Keplerian parameters are roughly Gaussian, then the 95\% confidence interval ($2\sigma$) should be roughly double our $1\sigma$ accuracy results. We find that for each parameter, this metric falls within $\pm 2\%$ of $2\sigma$. This is in line with the distributions of $\Delta$ values for each parameter shown in Figure \ref{fig:allparams}, which appear Gaussian.

Unsurprisingly, in both cases we see smaller (better) accuracy for retrievals using a Lambertian phase curve compared to retrievals done with no photometric information, or those using a Henyey-Greenstein phase function. We also see better accuracy in all three retrieval scenarios with 3 epochs versus 2, as expected. When the uncertainties on astrometry and photometry are larger and the planet is only detected twice, the accuracy is slightly worse than if no photometric information was included at all. Additionally, with 2 epochs of data and a Henyey-Greenstein phase function, $\sigma_\Delta$ is found to have a stronger dependence on the photometric uncertainty, given that the slope of the contours are steeper. When only astrometric uncertainty is higher, $\sigma_{a}$ is worse than when only photometric uncertainty is increased, with the exception of a planet detected twice with an unknown phase function. When a planet with a known scattering phase parameter is detected twice, increased astrometric uncertainty leads to a semi-major axis accuracy that is 1.84 times worse than if photometric uncertainty is increased. When a planet is detected three times with a known scattering phase function this metric is 1.74 times worse when astrometric uncertainty is increased versus when photometric uncertainty is increased. When the phase function is unknown, it is 1.33 times worse. When a planet with an unknown phase curve is detected twice, increased photometric uncertainty leads to accuracy that is 1.19 times worse than if astrometric uncertainty is increased. Extrapolating Figure \ref{fig:3epoch_contour} suggests that photometry stops being useful for orbit retrieval when photometric uncertainties are on the order of 10\% (corresponding to 10$\sigma$ detection of planetary flux).

We examine the effects of added photometric information on all Keplerian parameters in Figure \ref{fig:allparams} and Table \ref{tab:numbers}. We find that photometric information always improves parameter retrieval results with the exception of $\Omega$ and $M_o$ for retrievals done with 2 epochs. This is expected since these two parameters do not influence planet phase curves. 

While $a$ and $e$ are intrinsic to the planet's orbit, the other parameters are dependent on the observer's position. Retrievals of $a$ and $e$ are always most improved when the phase curve of the planet is Lambertian, and when it has been detected 3 times. Photometric information most significantly improves $\omega$, which makes sense as it describes the point of closest approach to the host star relative to its ascending node. It follows that information on a planet's changing brightness would significantly improve this parameter over simply using astrometric information.

\section{Discussion }\label{sec:conclu}

\subsection{Caveats}

We have assumed that the only constraints available are planetary astrometry and photometry. Our results are of merely academic interest if directly-imaged planets are discovered and their orbits retrieved via stellar astrometry \citep{meunier2022new}. Likewise, if the orbits are first characterized via stellar radial velocity \citep{Li_2021}, then the only parameter left to constrain is $\Omega$, which is immune to photometric constraints. 

By neglecting non-detections we have made the orbit retrieval problem somewhat easier: \cite{Guimond_2019} noted that a non-detection epoch provides less than half the astrometric information as a detection epoch, and there is no photometric information whatsoever.  On the other hand, by limiting out analysis to planets that are detected at all epochs, we have biased ourselves in favour of planets on face-on orbits, for which phase variations are more muted. We don't expect either of these biases to significantly impact our results because non-detections are relatively rare: 90\% of our synthetic single-epoch observations result in a detection, and 85\% result in detections at two epochs.  

In our numerical experiment we assume that the imaged planets are alone in their star system. The presence of other planets could change whether additional photometry improves results. If another planet is imaged while the target is within the IWA, its photometry could lead us astray. 

Regardless of the HG parameter $g$ chosen, our model assumes the planet has a spatially uniform albedo, or that integration times average over rotational variability.  \cite{Cowan_2012} showed that latitudinal albedo variations on a diffusely-reflecting planet can masquerade as specular reflections in the planet's orbital phase variations, i.e., $g<0$. A partially cloudy planet, on the other hand, might require a \emph{two term} Henyey-Greenstein phase function  \citep{Cahoy_2010,Feng_2018}. Since ours is the first study of photometry for direct imaging orbit retrieval, we limit ourselves to the single-parameter HG scattering phase function as a compromise between the standard Lambertian and the higher-order HG phase functions.

We also assume that images of a planet will be taken at fixed intervals of 90 days from the first epoch, which \cite{Guimond_2019} demonstrate to be near-optimal if the planet has an orbital period similar to Earth's. Depending on our knowledge of the phase curve \textit{a priori} this may not be the case. Rather, it could be preferable to image a planet at times where the brightness has changed most significantly.

We have limited our study to planets orbiting in the habitable zone of Sun-like stars. This choice was motivated by the fact that HabEx and LUVOIR would primarily target FGK stars to search for Earth twins. Indeed, yield estimates for these missions focused on Sun-like stars \citep{2014ApJ...795..122S,2019AAS...23340205S}. As discussed by \cite{Guimond_2019}, planets orbiting in the habitable zones of lower mass stars will more likely be obscured by the IWA, and if detected the optimal cadence for orbital retrieval may vary. In the context of reflected light, Earth analogs in the habitable zone of lower-mass stars will tend to have a more favourable contrast ratio due to their greater $R_{p}/a$.

\subsection{Summary}
Our results demonstrate a clear improvement in the retrieval of exoplanetary orbits given additional photometric information. We focused on the accuracy and precision of the retrieved semi-major axis since it is an intrinsic property of the orbit and first-order determinant of a planet's climate. 

We showed that if the scattering phase functions of exoplanets can be predicted a priori, then photometry provides a 50\% improvement in the efficiency of astrometric orbit retrieval: it takes three epochs of astrometry to constrain a planet's semi-major axis to 5\%, whereas the combination of astrometry and photometry achieves the same accuracy in only two epochs. For comparison, a single epoch of astrometry, combined with reasonable priors, constrains the semi-major axis to approximately 30\%.

If, on the other hand, we presume no prior knowledge of the scattering parameter $g$, then photometry only improves the two-epoch accuracy by $\sim$10\% but improves the three-epoch accuracy by a factor of 2. With four or more epochs of astrometry, the use of photometry only improves the retrievals by a few percent.

These results could indicate significant time and cost reduction for future direct imaging missions operating in visible light, such as HabEx or LUVOIR. We present a strong argument for the use of this additional photometric information if these missions anticipate operating with as few detections as possible. Constraining orbits more accurately and efficiently can improve estimates as to whether a planet is habitable, and whether it should be revisited for detailed characterization.  

Our fiducial case assumes precise astrometry and photometry. Larger uncertainties on either measurements reduce the benefits of photometry. We find that increasing photometric uncertainty has a less significant impact than increasing astrometric uncertainty. This indicates that in most cases any measurement of changing brightness could be beneficial to orbit retrievals. 

\subsection{Conclusions}
We have focused on Earth-like exoplanets orbiting Sun-like stars because they are the metric by which future direct imaging missions are compared \citep{2014ApJ...795..122S,stark2019exoearth,decadal}. However, the principles outlined here apply equally well to other exoplanets imaged in reflected light and may be useful for the Roman Space Telescope. 

The improved efficiency of using photometry is most useful for starshades, which slew much slower than a telescope equipped with a coronagraph.  Hence a direct-imaging survey with a starshade pays dearly for each additional epoch.  Even for a coronagraphic direct imaging campaign, the settling time after a slew can be comparable to the integration times, so reducing the number of revisits before triage of targets will improve the mission efficiency. 

Well-calibrated planetary photometry has uses beyond orbit determination.  At the very least, three or more epochs of astrometry plus photometry begin to uniquely constrain the HG phase curve parameter, hence hinting at the nature of the scattering mechanism \citep{henyey_greenstein_1941}. The shape of the phase curve may also betray latitudinal albedo variations \citep{Cowan_2012}.  Multi-band photometry, even at only 2--3 epochs, would strongly constrain the scattering properties of the planet; simultaneous multi-band photometry is easiest to envision with a starshade.

\section*{Acknowledgements}

This work was supported by the McGill Space Institute, TEPS CREATE and NSERC. We thank Ben Farr for his valuable help with \texttt{kombine}. We also thank fellow MEChA group members Lisa Dang, Samson Mercier, Alex Gass, and Taylor Bell for their help working through various hiccups along the way. We thank Chris Stark and Ell Bogat for useful conversations at a 2021 STSci Virtual Symposium. Finally, we thank the anonymous referee for their constructive review of the initial manuscript.

\section*{Data Availability Statement}

The data underlying this article will be shared on reasonable request to the corresponding author.




\bibliographystyle{mnras}
\bibliography{MyBibliography} 


\appendix
\section{Additional Data Table}\label{sec:a}
\begin{table*}
\begin{center}
\scalebox{1}{
    \begin{tabular}{c|c|c|c|c|c|c}
    \hline
    \hline
    Parameter & Epochs & & $\langle\Delta\rangle_{L}$ & $\sigma_{\Delta_{L}}$ &  $\langle\Delta\rangle_{HG}$ & $\sigma_{\Delta_{HG}}$ \\
    \hline
     & 2 & astro & -0.004 & 0.04 &  &  \\
     &  & astro + photo & -0.01 & 0.03 & -0.02 & 0.04 \\
    \cline{2-7}
     & 3 & astro & -0.004 & 0.04 &  &  \\
     $a$ &  & astro + photo & 0.001 & 0.01 & -0.0005 & 0.01 \\ 
    \cline{2-7} 
     & 4 & astro & 0.002 & 0.01 &  &  \\
     &  & astro + photo & 0.0001 & 0.007 & 0.0004 & 0.007 \\ 
    \hline
    \hline 
     & 2 & astro & -0.05 & 0.07 &  & \\
     &  & astro + photo & -0.01 & 0.03 & -0.02 & 0.05 \\ 
    \cline{2-7}
     & 3 & astro & -0.002 & 0.03 &  &  \\
    $e$ &  & astro + photo & -0.001 & 0.006 & -0.004 & 0.02 \\
    \cline{2-7} 
     & 4 & astro & -0.001 & 0.008 &  &  \\
     &  & astro + photo & -0.0006 & 0.004 & -0.0009 & 0.006 \\
    \hline 
    \hline 
     & 2 & astro & -0.003 & 0.07 &  &  \\
     &  & astro + photo & 0.008 & 0.03 & 0.008 & 0.03 \\
    \cline{2-7}
     & 3 & astro & -0.0003 & 0.02 &  &  \\ 
    $\cos i$ & & astro + photo & -0.0009 & 0.01 & -0.002 & 0.01 \\
    \cline{2-7} 
     & 4 & astro & -0.001 & 0.009 &  &  \\
     &  & astro + photo & -0.000006 & 0.007 & 0.0008 & 0.007 \\ 
    \hline 
    \hline 
     & 2 & astro & -0.007 & 0.2 &  &  \\ 
     &  & astro + photo & -0.007 & 0.04 & 0.005 & 0.07 \\ 
    \cline{2-7}
     & 3 & astro & -0.02 & 0.4 &  &  \\ 
    $\omega$ &  & astro + photo & 0.0007 & 0.02 & 0.004 & 0.03 \\ 
    \cline{2-7} 
     & 4 & astro & -0.007 & 0.07 &  &  \\
     &  & astro + photo & 0.0004 & 0.01 & -0.02 & 0.2 \\ 
    \hline 
    \hline 
     & 2 & astro & 0.02 & 0.3 &  &  \\
     &  & astro + photo & -0.005 & 0.3 & 0.03 & 0.3 \\ 
    \cline{2-7} 
     & 3 & astro & 0.03 & 0.6 &  &  \\ 
    $\Omega$ &  & astro + photo & 0.006 & 0.4 & -0.07 & 0.3 \\ 
    \cline{2-7}
     & 4 & astro & 0.05 & 0.3 &  &  \\ 
     &  & astro + photo & 0.01 & 0.2 & 0.03 & 0.3 \\ 
    \hline 
    \hline 
     & 2 & astro & 0.009 & 0.4 &  &  \\ 
     &  & astro + photo & 0.02 & 0.3 & -0.05 & 0.2 \\ 
    \cline{2-7}
     & 3 & astro & -0.01 & 0.4 &  & \\ 
    $M_o$ &  & astro + photo & -0.008 & 0.3 & 0.01 & 0.2 \\ 
    \cline{2-7} 
     & 4 & astro & -0.03 & 0.2 &  & \\ 
     &  & astro + photo & -0.016 & 0.25 & -0.01 & 0.2 \\ 
    \hline 
    \hline
    \end{tabular}}
    \caption{Bias, $\langle \Delta \rangle$, and accuracy, $\sigma_{\Delta}$, for 6 Keplerian parameters. The true Keplerian values for each planet are randomly generated following the input distributions outlined in Table \ref{table:params} in \S \ref{sec:synthetic}. These values are indicated for models with only astrometry, and then for models with astrometry + photometry with either a Lambertian or Henyey-Greenstein phase curve. These results are repeated for planets imaged at 2, 3, and 4 epochs.}
    \label{tab:numbers}
\end{center}
\end{table*}

\bsp	
\label{lastpage}
\end{document}